\documentclass[runningheads]{llncs}
\usepackage[T1]{fontenc}

\usepackage{graphicx}
\usepackage{cleveref}
\usepackage[table,xcdraw]{xcolor}
\usepackage{array}
\newcolumntype{x}[1]{>{\centering\let\newline\\\arraybackslash\hspace{0pt}}p{#1}}

\usepackage{placeins}
\begin{document}
\title{Demonstrating \emph{ARG-V}'s Generation of Realistic Java Benchmarks for SV-COMP }

\author{Charles Moloney\inst{1}\orcidID{0009-0005-0589-5389}
\and Robert Dyer\inst{1}\orcidID{0000-0001-9571-5567}
\and Elena Sherman\inst{2}\orcidID{0000-0003-4522-9725}}
\authorrunning{Moloney et al.}
\titlerunning{Demonstrating \emph{ARG-V}}

\institute{University of Nebraska--Lincoln, Nebraska, USA\\
\email{\{cmoloney2@huskers.unl,rdyer@unl\}.edu} 
\and 
Boise State University, Idaho, USA\\
\email{elenasherman@boisestate.edu}}

\maketitle

\begin{abstract}
The SV-COMP competition provides a state-of-the-art platform for evaluating software verification tools on a standardized set of verification tasks. Consequently, verifier development outcomes are influenced by the composition of program benchmarks included in SV-COMP. When expanding this benchmark corpus, it is crucial to consider whether newly added programs cause verifiers to exhibit behavior distinct from that observed on existing benchmarks. Doing so helps mitigate external threats to the validity of the competition’s results.

~~\quad In this paper, we present the application of the \emph{ARG-V} tool for automatically generating Java verification benchmarks in the SV-COMP format. We demonstrate that, on a newly generated set of 68 realistic benchmarks, all four leading Java verifiers decrease in accuracy and recall compared to their performance on the existing benchmark suite. These findings highlight the potential of \emph{ARG-V} to enhance the comprehensiveness and realism of verification tool evaluation, while also providing a roadmap for verifier developers aiming to improve their tools’ applicability to real-world software.

\end{abstract}

\section{Introduction}
One of the persistent challenges in developing software verifiers is a lack of publicly available and representative benchmark programs for evaluating functionality and identifying new edge cases to be solved~\cite{benchmark-challenges}. To address this problem, researchers have come together to share resources and provide a common platform for benchmarking in groups like SV-COMP~\cite{SVCOMP25}. These have been a massive boon to the formal verification community, but it is important to be aware of how the nature of these benchmarks can influence competition results and, ultimately, verifier development. To be considered as valid Java SV-COMP benchmarks, programs must only leverage the Java Class Library, inject a shared ``Verifier'' library, and have an associated property key for evaluation. Generation of a realistic (i.e., mined from real-world code) SV-COMP Java benchmark would typically consist of: 

\begin{enumerate}
    \item Identification of code with desirable characteristics, such as conditional statements or arithmetic expressions.
    \item Removal of non-JDK dependencies and irrelevant code fragments.
    \item Instrumenting programs with a common ``Verifier'' variable(s).
    \item Creation of verification scaffolding files with expected verdicts.
    \item Definition of properties for verification tasks and modification of code to target them.
\end{enumerate}

Performing each step manually is labor-intensive and requires human intervention, which might introduce bias; benchmark selection alone can lead to a disproportionate number of benchmarks exhibiting characteristics that verifiers can already efficiently handle. Impartial selection would reduce the number of opportunities for bias and increase the likelihood of new benchmarks exploring underdeveloped features for existing verifiers. In an effort to mitigate this bias and enable developers to skip directly to the final step of the generation process, we utilize \emph{ARG-V}~\cite{argv:website}, our tool for automating the time-consuming filtering and formatting necessary to get a benchmark into a required format. 

This paper focuses on the application of \emph{ARG-V} to meaningfully augment the SV-COMP corpus. It explores how the \emph{ARG-V} generated benchmarks increase the representation of program behaviors, which reduces threats to the external validity of SV-COMP results. We have generated a new set of benchmarks for consideration at SV-COMP and evaluated on them the behavior of four prominent Java verifiers. At the time of this paper, 70 \emph{ARG-V} benchmarks have been merged into the SV-COMP Java library across the categories of reachability safety and runtime exception safety. By leveraging the impartiality and filter functionality of our tool, we hope to assist the formal verification community by accelerating the production of more benchmarks, pushing the boundaries and coverage to new areas in existing public corpora.

\section{\emph{ARG-V} Tool Description}

\emph{ARG-V} builds on a previous tool, \emph{PAClab: A Program Analysis ColLABoratory}~\cite{paclab}, by expanding its capabilities and offering the option to perform transformations targeted to the SV-COMP format. While \emph{PAClab} is designed to create benchmarks on the intra-procedural level, transforming methods using symbolic booleans and integers for Java Symbolic PathFinder~\cite{SPF}, \emph{ARG-V} substantially extends \emph{PAClab}'s scope to handle artifact preparation, additional data types, field variables, and cross-class interactions. \Cref{tab:comparison} highlights the major changes between the tools. 

\begin{table}[t]
\caption{Comparison showing the expanded functionality of ARG-V}
\label{tab:comparison}
\Large
\resizebox{\textwidth}{!}{%
\rowcolors{1}{gray!20}{white}
\begin{tabular}{l|l|l}
\textbf{Feature}     & \textbf{PAClab}                                                               & \textbf{ARG-V}                                                                                                                                                     \\
Scope                & \begin{tabular}[c]{@{}l@{}}Intra-procedural\\ (Method-level only)\end{tabular} & \begin{tabular}[c]{@{}l@{}}Inter-procedural\\ (Class-level and JDK packages)\end{tabular}                                                                           \\
Member Variables     & Static only                                                                   & Static and Non-Static                                                                                                                                              \\
Supported types      & Only ints and booleans                                                        & All primitive types and primitive arrays                                                                                                                                                \\
Type Resolution      & Implicit resolution only~                                                      & \begin{tabular}[c]{@{}l@{}}Implicit and Explicit resolution via method bindings\\ in classpath and type table\end{tabular}                                         \\
Artifact Preparation~ & None                                                                          & \begin{tabular}[c]{@{}l@{}}Creation of main method and benchmark config (yml) file, \\  preservation of provenance information, and re-naming \\ to formatting standard\end{tabular}
\end{tabular}%
}
\end{table}

\emph{ARG-V}'s architecture consists of three distinct modules:
\begin{enumerate}
    \item \textbf{Download}, which fetches a number of GitHub repositories as specified by a CSV input file.
    \item \textbf{Filter}, which narrows down the repository files to those containing user-set minimums of specified features, such as if statements, type expressions, type parameters, or conditionals.
    \item \textbf{Transform}, which breaks the code down into an abstract syntax tree (AST) and performs a series of modifying operations, preparing the file for use as a benchmark and generating any necessary artifacts
\end{enumerate}

The transform module leverages the Eclipse Foundation's JDT~\cite{eclipse_jdt} for AST traversals, using a type table and method bindings to make informed decisions about code removal and replacement (a more detailed explanation is described in the original \emph{PAClab} paper~\cite{paclab}). This allows the code to compile independently of its package or source repository without losing all meaning, strategically removing external dependencies while keeping as much computational logic as possible.

\section{Generation and Testing of New Benchmarks}
To demonstrate the efficacy of \emph{ARG-V} as a benchmark generator, we use it to generate 68 new benchmarks to augment the SV-COMP Java corpus~\cite{replication-package}. These benchmarks encompass two categories: \texttt{ReachSafety}, which resolves to true if all assertions hold in the code, and \texttt{ExceptionProperty}, which resolves to true if the code cannot throw a RuntimeException. Some of these benchmarks contain noteworthy characteristics for both categories, so the resulting set consists of 48 \texttt{ReachSafety} benchmarks (31 true, 17 false) and 50 \texttt{ExceptionProperty} benchmarks (39 true, 11 false) -- 98 property runs in total. 

For the generation of these benchmarks, \emph{ARG-V} scrapes the first 1,261 repositories within the RepoReaper database~\cite{reporeaper}, an open-source curator of publicly available GitHub repositories. We set the criteria to require at least one `if` statement that is conditional on a chosen primitive type (floating point or integer) to ensure multiple paths through the code.  This minimal criterion guarantees that files without interesting logic (e.g., only getters and setters) will not be downloaded.  We avoid stricter criteria that would miss useful benchmarks, as they result in a less impartial result set by concentrating the resulting corpus around a very specific criterion, such as 3 or more branching conditions. Assert statements are added in manually, checking for variable value ranges or null-safety. During any manual modifications, an emphasis is put on basic constructs, primitive values, and arrays, with no recursion\footnote{From our experience, we find that recursion often leads to timeouts or unknown verdicts from verifiers.  The decision was made to avoid recursive benchmarks when generating our sample corpus, as it might not lead to meaningful results from the verifiers on our new benchmarks.} or complex data structures.  The final benchmarks contain an average of 41.7 lines of code per file, comparable to the average 42.2 lines of code in existing SV-COMP Java benchmarks~\cite{svbenchmarks2025}. 

The main goal of this tool demonstration is to assess whether \emph{ARG-V} is capable of producing a distinct corpus of benchmarks that provide a fresh challenge for Java verifiers, as well as whether there is a significant difference in difficulty between the realistic \emph{ARG-V} benchmarks and the existing SV-COMP corpus. As a representation of the state of the art in Java verification, the new benchmarks are evaluated using the four top-performing Java verifiers from SV-COMP 2025~\cite{SVCOMP25}: MLB~\cite{mlb_paper,mlb_artifact}, GDart~\cite{gdart_paper,gdart_artifact}, JavaRanger~\cite{javaranger_paper,javaranger_artifact}, and JBMC~\cite{jbmc_paper,jbmc_artifact}.

To standardize our evaluation, experiments are run using SoSy-Lab's \\
BenchExec~\cite{benchexec_paper}, the same platform used in the SV-COMP competition. Each tool is executed on a verifier run and limited to 4GB of memory, 120 seconds of cumulative CPU time, and 2 processing units. The benchmarking machine is an Intel Core i9-12900HK CPU at 2.5GHz, 32 GB of RAM, running Ubuntu 24.04.1 LTS with BenchExec version 3.30~\cite{benchexec_release}. All benchmarks are run using the same parameters; the complete command for executing JBMC can be seen below:

\begin{small}
\begin{verbatim}
    scripts/execute_runs/execute-runcollection.sh \
    benchexec/bin/benchexec   archives/2025/jbmc.zip  \
    benchmark-defs/jbmc.xml   witness.graphml  results-verified/ \
    -t ReachSafety-Java -t RuntimeException-Java \
    --timelimit 120 --memorylimit 4GB --limitCores 2 
    --numOfThreads 8 --read-only-dir / --overlay-dir /home/ \
    --overlay-dir ./ --tool-directory . -o .graphml
\end{verbatim}
\end{small}

\section{Evaluation of \emph{ARG-V} Benchmarks against SV-COMP}

The same verifiers are also run with the same parameters on the existing SV-COMP corpus of 674 Java benchmarks, which contains of 674 \texttt{ReachSafety} properties and 673 \texttt{ExceptionProperty} properties. We analyze the differences in verification results between these two benchmark sets. To avoid overlap, the SV-COMP benchmarks produced by \emph{ARG-V} are omitted from this corpus.

We compare the results of the verifiers on the two benchmark corpora using standard classification metrics: \textbf{accuracy}, the proportion of correctly classified samples; \textbf{precision}, the fraction of predicted positives that are actually true positives; \textbf{recall}, the fraction of actual positives correctly identified versus the total number of actual positives; and \textbf{specificity}, the fraction of actual negatives that were correctly identified compared to the total number of actual negatives. 

These values are calculated for both categories, as well as the cumulative score across \texttt{ReachSafety} and \texttt{ExceptionProperty}. Aggregate results from all verifiers are available in \Cref{tab:excluding-unknown}; complete results and raw data are available in the replication package~\cite{replication-package}. Results for \Cref{tab:excluding-unknown} are calculated strictly on affirmative outputs, thus omitting undecidable results, i.e., when a verifier returns  ``unknown'', ``error'', or ``timeout'' rather than ``true'' or ``false''. 

Across all reach-safety and cumulative metrics, we find that verifiers' performances \textbf{decrease substantially} on the new, realistic benchmarks compared to the existing SV-COMP corpus. When evaluating if a program will ever have false assertions, the verifiers are less than half as likely to make a correct decision. The verifiers do perform relatively consistently between the new and existing benchmarks within the runtime-exception category, but only when undecidable results are omitted from the results.

\begin{table}[t]
\centering
\caption{Classification metrics between benchmark sets, excluding ``unknown'', ``error'', and ``timeout''}
\label{tab:excluding-unknown}
\begin{tabular}{|c|x{1.5cm}|x{1.5cm}|x{1.5cm}|x{1.5cm}|x{1.5cm}|x{1.5cm}|}
\cline{2-7}
\multicolumn{1}{c|}{~} & \multicolumn{2}{x{3cm}|}{\textbf{reach-safety}} & \multicolumn{2}{x{3.1cm}|}{\textbf{runtime-exception}} & \multicolumn{2}{x{3cm}|}{\textbf{cumulative}} \\
\cline{2-7} 
\multicolumn{1}{c|}{~} & \textbf{existing} & \textbf{new} & \textbf{existing} & \textbf{new} & \textbf{existing} & \textbf{new} \\
\hline
Accuracy    & 1.00 & 0.72 & 0.67 & 0.72 & 0.83 & 0.72 \\
\hline
Precision   & 0.99 & 0.96 & 1.00 & 0.89 & 0.99 & 0.90 \\
\hline
Recall      & 1.00 & 0.49 & 0.66 & 0.73 & 0.75 & 0.64 \\
\hline
Specificity & 0.99 & 0.98 & 0.91 & 0.68 & 0.99 & 0.88 \\
\hline
\end{tabular}
\end{table}

SV-COMP also allows verifiers to return an ``unknown'' result when they cannot solve a verification task~\cite{SVCOMP25}. It is relevant to analyze the performance on these benchmarks while including this uncertainty option, as a higher inability to make a judgment on a new set of benchmarks suggests potential categories of benchmarks that the verifier is not well-equipped to handle. \Cref{tab:with-unknown} shows the same classification metrics as before, but with undecidable metrics counting as a failed result. Precision is omitted, as there is no unknown-inclusive analog for this metric.

With undecidable results included, the verifiers have a \textbf{notably poorer performance} for all metrics across both checked properties. The runtime-exception property, which verifiers exhibit comparable performance on for unknown-exclusive data, has a much larger disparity in \Cref{tab:with-unknown}. This is due to a substantially higher proportion of undecidable responses on the new benchmarks.

\begin{table}[t]
\centering
\caption{Undecidable-Inclusive metrics with ``unknown'', ``error'', or ``timeout'' counting as a failed result. Precision is unaffected by this change.}
\label{tab:with-unknown}
\begin{tabular}{|c|x{1.5cm}|x{1.5cm}|x{1.5cm}|x{1.5cm}|x{1.5cm}|x{1.5cm}|}
\cline{2-7}
\multicolumn{1}{c|}{~} & \multicolumn{2}{x{3cm}|}{\textbf{reach-safety}} & \multicolumn{2}{x{3.1cm}|}{\textbf{runtime-exception}} & \multicolumn{2}{x{3cm}|}{\textbf{cumulative}} \\
\cline{2-7} 
\multicolumn{1}{c|}{~} & \textbf{existing} & \textbf{new} & \textbf{existing} & \textbf{new} & \textbf{existing} & \textbf{new} \\
\hline
Accuracy    & 0.75 & 0.33 & 0.49 & 0.35 & 0.62 & 0.34 \\
\hline
Recall      & 0.71 & 0.18 & 0.49 & 0.35 & 0.55 & 0.27 \\
\hline
Specificity & 0.76 & 0.60 & 0.54 & 0.34 & 0.75 & 0.50 \\
\hline
\% Undecidable  & 25\% & 55\% & 26\% & 52\% & 26\% & 53\% \\
\hline
\end{tabular}
\end{table}

These results suggest a significant difference in verifier capability between the two corpora. A cumulative recall of only 0.271 for the new benchmarks versus the 0.554 of the SV-COMP corpus means verifiers are less than half as likely to be able to correctly identify a property as true on a similarly-scoped benchmark when it does not come from the SV-COMP suite. A substantial drop in specificity denotes a similar inability to correctly identify a property as false, resulting in an overall much lower accuracy on new benchmarks.

\Cref{tab:individual-verifiers} breaks the results out across individual verifiers. For instance, results for GDart~\cite{gdart_paper,gdart_artifact} indicate it is possible the model is overfitting on the current SV-COMP corpus, given its much lower performance and higher unknown/timeout/error rate on new benchmarks of similar complexity.

\begin{table}[t]
\centering
\caption{Results of individual verifiers across all metrics. Undecidable Inclusive (UI) metrics include ``unknown'', ``error'', or ``timeout'' as a failed result.}
\label{tab:individual-verifiers}
\resizebox{\columnwidth}{!}{%
\begin{tabular}{|c|x{1.4cm}|x{1.4cm}|x{1.4cm}|x{1.4cm}|x{1.4cm}|x{1.4cm}|x{1.4cm}|x{1.4cm}|}
\cline{2-9}
\multicolumn{1}{c|}{~} & \multicolumn{2}{x{2.8cm}|}{\textbf{java-ranger}} & \multicolumn{2}{x{2.8cm}|}{\textbf{mlb}} & \multicolumn{2}{x{2.8cm}|}{\textbf{gdart}} & \multicolumn{2}{x{2.8cm}|}{\textbf{jbmc}} \\
\cline{2-9} 
\multicolumn{1}{c|}{~} & \textbf{existing} & \textbf{new} & \textbf{existing} & \textbf{new} & \textbf{existing} & \textbf{new} & \textbf{existing} & \textbf{new} \\
\hline
Accuracy       & 0.70 & 0.67 & 0.66 & 0.51 & 1.00 & 0.86 & 0.99 & 0.83 \\
\hline
Precision      & 0.99 & 1.00 & 0.98 & 0.74 & 1.00 & 1.00 & 1.00 & 0.95 \\
\hline
Recall         & 0.57 & 0.50 & 0.49 & 0.43 & 1.00 & 0.75 & 0.99 & 0.78 \\
\hline
Specificity    & 0.99 & 1.00 & 0.99 & 0.68 & 1.00 & 1.00 & 1.00 & 0.92 \\
\hline
\rowcolor[HTML]{EFEFEF} 
UI-Accuracy    & 0.48 & 0.08 & 0.56 & 0.31 & 0.63 & 0.32 & 0.81 & 0.64 \\
\hline
\rowcolor[HTML]{EFEFEF} 
UI-Recall      & 0.40 & 0.06 & 0.39 & 0.24 & 0.59 & 0.21 & 0.84 & 0.57 \\
\hline
\rowcolor[HTML]{EFEFEF} 
UI-Specificity & 0.65 & 0.14 & 0.91 & 0.46 & 0.71 & 0.57 & 0.76 & 0.82 \\
\hline
\rowcolor[HTML]{EFEFEF} 
\% Undecidable     & 32\% & 88\% & 18\% & 22\% & 37\% & 63\% & 18\% & 22\% \\
\hline
\end{tabular}}
\end{table}

The ability of \emph{ARG-V} to na\"ively (i.e., no deliberate attempt to fool verifiers or intentional selection of arbitrarily complex code) produce benchmarks that are difficult for the current state-of-the-art verifiers is promising and shows the importance of a richer set of benchmark programs obtained from open-source code. We intend for this tool to be used to extend the existing boundaries of software verifiers further, and doing so will require benchmark suites that are more comprehensive and explore new, unaddressed areas.

Leveraging the parameterized filter step, users can further test for specific features of interest. For instance, many of the individual benchmarks that see a high failure rate within the \emph{ARG-V} sample corpus involve multiple branching conditions operating on nondeterministic floating point values~\cite{replication-package}; a researcher hoping to test and identify more failing edge cases for this property can filter for benchmark candidates, using parameters such as a high \texttt{minIfStmt} threshold, and select methods using symbolic \texttt{double} and \texttt{float} types.

\section{Conclusion and Future Work}
We present \emph{ARG-V}, an Adaptable Realistic Benchmark Generator for Verification. Using this tool, we generate a sample corpus of 68 benchmarks for use in the SV-COMP Java benchmark library, and these benchmarks are capable of providing a substantial increase in challenge for the state of the art in software verification, suggesting unknown blind spots in verifier logic and/or overfitting to public benchmark repositories. We believe this tool provides an efficient and streamlined method for generating new benchmarks at scale, illuminating new unaddressed problem spaces, and identifying more edge cases for verifiers. The underlying source(s) for the higher difficulty solving tests generated by \emph{ARG-V} remains a largely open question. We can confidently say it is not caused by the length of the code (as the \emph{ARG-V} generated benchmarks are shorter), the use of floating-point operations (as verifiers do well on prior benchmarks with floating-point), or the use of Java Standard Library functions.  We hope to use this set of benchmarks as a starting point to investigate this specific problem. Additionally, we will continue to work to make development for verifier researchers faster and more comprehensive, and we plan to continue to create more benchmarks using \emph{ARG-V} as we add new features. This tool can still be expanded in a variety of ways: we are actively working to add more granular filtering methods, supporting the transformation of more data structures and edge cases within the Java library, and exploring leveraging deep learning to prevent creating benchmarks already covered by similar tests within a corpus. Development is also underway for an analogous C-code transformer~\cite{argc-transformer}.

\begin{credits}
\subsubsection{\ackname} We thank Jon Craig and Rubal Goyal for helping with an earlier implementation of the tool and benchmarks. This tool was partially funded by the U.S. National Science Foundation under grants 1823294, 1823357, 2346326, and 2346327.
\end{credits}

\section*{Data Availability}
The source code and build steps are available in the replication package~\cite{replication-package}.  For reproduction, use version 1.0.1, and for general reuse, use the most recent version available.

\bibliographystyle{splncs04}
\bibliography{refs}

\end{document}